\documentclass[
reprint,
superscriptaddress,
showpacs,
 amsmath,amssymb,
floatfix,
]{revtex4-1}
\usepackage{subcaption}
\usepackage{dcolumn}
\usepackage{bm}
\usepackage{turnstile}

\usepackage{graphicx}
\usepackage{amsmath,amssymb}
\usepackage{url}

\usepackage{epsfig}
\usepackage{epstopdf}

\begin{document}

\title{
Proximity effect in [Nb(1.5nm)/Fe(x)]$_{10}$/Nb(50nm) superconducting/ferromagnet heterostructures
}
\author{Yu.~N.~Khaydukov}
\affiliation{Max-Planck-Institut f\"ur Festk\"orperforschung, Heisenbergstra\ss e 1, D-70569 Stuttgart, Germany}
\affiliation{Max Planck Society Outstation at the Heinz Maier-Leibnitz Zentrum (MLZ), D-85748 Garching, Germany}
\affiliation{Skobeltsyn Institute of Nuclear Physics, Moscow State University, Moscow 119991, Russia}

\author{S.~P\"utter}
\affiliation{Forschungszentrum J{\"u}lich GmbH, J{\"u}lich Centre for Neutron Science (JCNS) at Heinz Maier-Leibnitz Zentrum (MLZ), Lichtenbergstr.\ 1, D-85747 Garching, Germany}

\author{L.~Guasco}
\affiliation{Max-Planck-Institut f\"ur Festk\"orperforschung, Heisenbergstra\ss e 1, D-70569 Stuttgart, Germany}
\affiliation{Max Planck Society Outstation at the Heinz Maier-Leibnitz Zentrum (MLZ), D-85748 Garching, Germany}
\author{R.~Morari}
\affiliation{Institute of Electronic Engineering and Nanotechnologies ASM, MD2028 Kishinev, Moldova}
\author{G.~Kim}
\affiliation{Max-Planck-Institut f\"ur Festk\"orperforschung, Heisenbergstra\ss e 1, D-70569 Stuttgart, Germany}
\author{T.~Keller}
\affiliation{Max-Planck-Institut f\"ur Festk\"orperforschung, Heisenbergstra\ss e 1, D-70569 Stuttgart, Germany}
\affiliation{Max Planck Society Outstation at the Heinz Maier-Leibnitz Zentrum (MLZ), D-85748 Garching, Germany}

\author{A.~S.~Sidorenko}
\affiliation{Institute of Electronic Engineering and Nanotechnologies ASM, MD2028 Kishinev, Moldova}

\author{B.~Keimer}
\affiliation{Max-Planck-Institut f\"ur Festk\"orperforschung, Heisenbergstra\ss e 1, D-70569 Stuttgart, Germany}
\date{\today}

\begin{abstract}
We have investigated the structural, magnetic and superconducting properties of [Nb(1.5nm)/Fe(x)]$_{10}$ superlattices deposited on a thick Nb(50nm) layer.  Our investigation showed that the Nb(50nm) layer grows epitaxially at 800$^\circ$C on Al$_2$O$_3$(1$\bar{1}$02) substrate. Samples grown at this condition posses a high residual resistivity ratio of 15-20. By using neutron reflectometry we show that Fe/Nb superlattices with $x<$ 4 nm form a depth-modulated FeNb alloy with concentration of iron varying between 60\% and 90\%. This alloy has properties of a weak ferromagnet. Proximity of this weak ferromagnetic layer to a thick superconductor leads to an intermediate phase that is characterized by co-existing superconducting and normal-state domains. By increasing the thickness of the Fe layer to $x$ = 4 nm the intermediate phase disappears. We attribute the intermediate state to proximity induced non-homogeneous superconductivity in the periodic Fe/Nb structure.
\end{abstract}


\maketitle

\section{Introduction}
The term proximity effect was first introduced in the 1960s when considering contact of a normal (N) metal and a superconductor (S). It was shown that the superconducting correlations can penetrate into the normal metal, so that the order parameter in the latter $\Psi(z) \sim exp(-z / \xi_N)$ is nonzero over a depth  $\xi_N$, which can reach values of several microns.  In ferromagnetic (F) materials the proximity effect exists also, however with one essential difference: the coherence length in the ferromagnet $\xi_F$ is complex, so that the order parameter is not only a damped but also an oscillatory function. In the most common case of dirty ferromagnets the penetration length can be expressed as $\xi_F = \sqrt{\hbar D_F/E_{ex}}$. Here $D_F$ is the diffusion constant, and $E_{ex}$  is the exchange field of the ferromagnet. In mean field theory \cite{Kittel} $E_{ex} \sim T_m$ where $T_m$ is the Curie temperature.

The proximity effect in S/F structures has great technological importance for creation  of spintronic devices where the transport properties of the structure are controlled via manipulation with magnetic order in the F subsystem \cite{soloviev2017beyond,golubov2017superconductivity,klenov19}. The best performance of such devices is realized for the case $d_F \sim \xi_F$. Strong ferromagnets like Fe, Co or Ni with  $E_{ex} \sim$ 1000K have typical $\xi_F$ of order of 1 nm \cite{SidorenkoAnnPhys,MuegePRL96,MuegePRB97,Nagy_2016}, which makes it difficult to create a homogeneous F layer of such small thickness. The natural way to increase $\xi_F$ is reduction of $T_m$, which can be accomplished e.g. by alloying strong ferromagnet with non-magnetic atoms. We can mention  copper-nickel \cite{ZdravkovPRL,Zdravkov.PRB,KhaydukovJAP15} and palladium-iron ferromagnets \cite{Glick17,Petrov19,Esmaeili19} ferromagnets where such alloying led to significant  suppression of $T_m$.

Iron and niobium is another pair of metals which have been intensively studied \cite{RehmPhysB96,MuegePRL96,RehmEPL97,KlosePRL97,MuegePRB97,MuegePhysC98,HuangJMMM06,Marczynska2018,Wachowiak18}. Iron is a strong ferromagnet with $T_m$=1044K and Nb is a superconductor with bulk superconducting transition temperature $T_C = 9.3K$. Proximity effects in Fe/Nb systems were extensively studied before \cite{MuegePRL96,MuegePhysC98,MuegePRB97,HuangJMMM06}. The RKKY coupling of Fe layers through Nb(y),  y=(1.3+0.9$\times n$)nm (n = 0,1,2) was reported in Refs. \cite{RehmPhysB96,RehmEPL97,KlosePRL97}. Moreover in \cite{KlosePRL97} it was shown that hydrogen uptake can influence the exchange coupling.

A peculiarity of Fe/Nb heterostructures is the high mutual solubility of Fe and Nb, leading to the formation of an FeNb alloy on the interface. In this work we aim to elucidate the structural, magnetic and superconducting properties of Fe/Nb superlattices in proximity to a thick superconducting Nb layer. To study the effect of alloying we varied the growth temperature as well as the Fe layer thickness.

\section{Experimental}
\subsection{Growth conditions and techniques description }
Samples of nominal structure Pt(3nm)/[Nb(1.5nm)/Fe(x)]$_{10}$/Nb(50nm) were prepared on Al$_2$O$_3$(1$\bar{1}$02) substrates using a DCA M600 MBE system with a base pressure of 10$^{-10}$ mbar. Before deposition, the substrates were cleaned from organic contaminations with ethanol  and isopropanol ex-situ and heated at 1000$^\circ$C in ultra high vacuum for 2-3 hours. A 50 nm thick Nb layer was deposited at a typical rate of 0.6\AA/s and substrate temperature $T_{Nb}$ = 800$^\circ$C for samples s1 to s5 and $T_{Nb}$ = 33$^\circ$C for sample s6. Subsequently, the substrate temperature was decreased to $T_{PS}$ = 30$^\circ$C-100$^\circ$C (see table \ref{Tab1}) and a periodic structure [Nb(1.5nm)/Fe(x)]$_{10}$ was deposited starting from the iron layer. The growth rates for both elements in the periodic structure were about 0.1 \AA/s. On top a 3 nm Pt cap layer was grown to protect the sample against oxidation at about 0.3 \AA/s at room temperature. Fe was deposited by thermal evaporation from an effusion cell while Nb and Pt were grown by electron beam evaporation. Reflection high energy  electron  diffraction  (RHEED) was measured in-situ during deposition to trace the structure of the atomic layer being deposited. For the RHEED experiment electron beam of 15 keV energy was directed along the [$20\bar{2}\bar{1}$] azimuth of the sapphire substrate.

\begin{figure}[htb]
\centering
\includegraphics[width=\columnwidth]{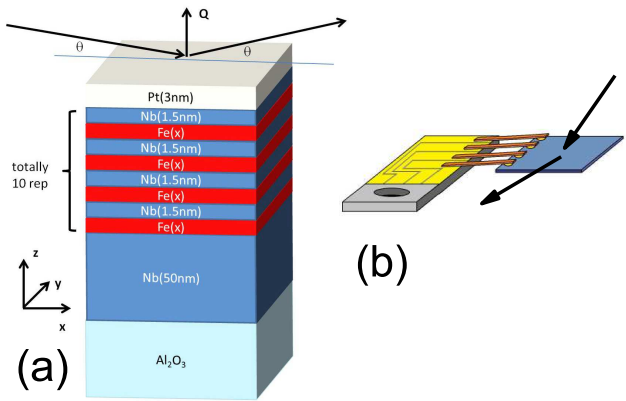}
\caption{
Sketch of the structure and reflectometric experiment (a) and set up for the transport measurements (b).
Black arrows show the direction of the neutron beam.
 }
\label{Fig1}
\end{figure}

In order to check the crystal structure and the quality of the epitaxial growth, X-ray diffraction measurements were performed using a $\theta$-2$\theta$ diffractometer. The diffractometer operates at the wavelength $\lambda$=1.54 \AA and is equipped with a DECTRIS line detector, which allowed simultaneous measurement of both specular and off-specular reflections.

The polarized neutron reflectometry (PNR) experiments were conducted on the angle-dispersive reflectometer NREX ($\lambda$ = 4.28 \AA) at the research reactor FRM-II (Garching, Germany). During the experiments we applied a magnetic field in-plane and normal to the sample plane. Data were fitted to models using the exact solutions of the Schr\"odinger equation as described in our prior work \cite{KhaydukovPRB14,KhaydukovPRB17,KhaydukovPRB19}.

For the transport experiment we used the device depicted in Fig. \ref{Fig1}b. The device consists of four metallic springs touching surface of the sample. The tension of the springs is high enough to ensure good contact with the sample surface and measure resistivity using a standard four-point contact method. The setup is designed to enable simultaneous PNR and transport experiments, though in this work we used it ex-situ. For the measurements we used an ac-current with an amplitude of $\sim$ 100-200 $\mu$A.  In the experiment we measured the resistance of samples $W$ as a function of temperature $T$ and magnetic field $H$ which was applied parallel to the sample surface. Before every $H$-scan we waited 10-15 minutes to stabilize the temperature. From the transport measurements we derived the residual resistivity ratio $RRR$ = $W$(300K)/$W$(10K), the superconducting transition temperature $T_C$ and its width  $\Delta T_C$. The latter two parameters were defined as the center and the width of derivative d$W$/d$T$, respectively.

\subsection{Reflectometry data analysis}
To study interdiffusion processes in a layered structure a reflectometry technique (X-ray or neutron) can be used \cite{Merkel10,Gong17,Merkel19}. In this method a reflectivity curve $R(Q)$ is measured as a function of momentum transfer $Q$ = $4\pi \sin(\theta) / \lambda$. In the kinematical approximation the reflectivity is proportional to the square of the Fourier transform of $d\rho(z) / dz$, where $\rho(z)$ is the depth profile of the scattering length density (SLD). The SLD is defined as the product of the averaged scattering length $\bar{b}$ and the density $N$. For a periodic structure with period $D$ repeated $n$-times one can write a simple expression for the reflectivity: \cite{AndreevaPhysRevB.72.125422}

\begin{equation}\label{Refl}
R(Q) = |L_n(Q,D) F(Q,\rho)|^2,
\end{equation}
where $L_n(Q,D)=(1-e^{inQD})/(1-e^{iQD})$ is the Laue function and $F(Q,\rho)$ - is the structure factor of the unit cell. The latter can be written for the case of Fe(x)/Nb(y) periodic bilayer as

\begin{equation}\label{StrucFac}
F(Q,\rho) = \frac{4\pi \Delta \rho}{Q^2} e^{iQx}(1-e^{iQy}),
\end{equation}
where $\Delta \rho = (\rho_{Fe}- \rho_{Nb})$ - is the contrast between the SLDs of Fe and Nb.  Thus from \eqref{Refl} and \eqref{StrucFac} it follows that reflectometry measures the contrast between SLDs of neighboring layers. Using \eqref{Refl} and \eqref{StrucFac} we can write for the reflectivity $R_1$ at the first Bragg peak $Q_1=2\pi/D$:

\begin{equation}\label{Contrast}
\Delta \rho = \frac{\sqrt{R_1} Q_1^2} {8 \pi n}.
\end{equation}

Thus the Bragg analysis allows us to determine the contrast between the SLDs of Fe and Nb. Interdiffusion will lead to suppression of the contrast and hence of $R_1$. Under assumption that the packing density $N_{av}$ is the same for both layers we may estimate the concentration of Fe in the Fe$_c$Nb$_{1-c}$ alloy as

\begin{equation}\label{Concent}
c = \frac{\bar{b}-b_{Nb}} {b_{Fe}-b_{Nb}},
\end{equation}
where $\bar{b} = \rho /N_{av}$ is the averaged coherent scattering length of a corresponding layer, and $b_{Fe}$ and $b_{Nb}$ are the coherent scattering lengths of Fe and Nb. The X-ray SLDs of Fe and Nb differ only by a few percent which makes the X-ray contrast very small even without interdiffusion. For neutrons, in contrast, the SLDs of Fe and Nb, $\rho_{Fe} = 8 \times 10^{-4}$ nm$^{-2}$ and $\rho_{Nb} = 3.9 \times 10^{-4}$ nm$^{-2}$, differ by a factor of two, which makes neutron reflectometry a better choice to study diffusion in periodic Fe/Nb structures. Another advantage of neutron reflectometry is its sensitivity to the magnetic depth profile. The total SLD for spin-up(+) and spin-down(-) neutrons can be written as  $\rho^\pm(z) = \rho_0(z) \pm \rho_m(z)$, where $\rho_0$ and $\rho_m$ are the nuclear and magnetic SLDs. The latter is proportional to the magnetization of a layer. Thus in addition to the chemical diffusion we can study "magnetic" diffusion.

 \subsection{Structural study}
  \subsubsection{Growth analysis with RHEED}
\begin{figure}[htb]
\centering
\includegraphics[width=1\columnwidth]{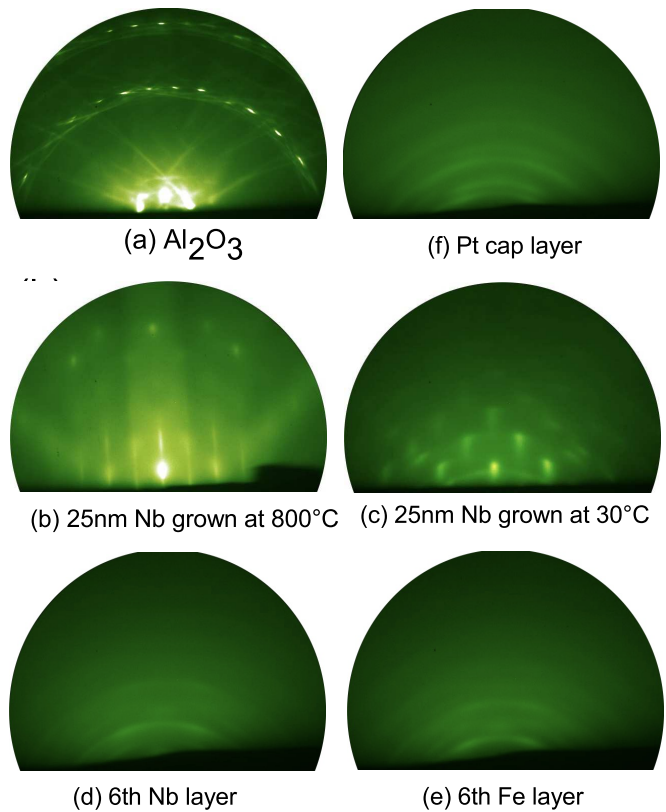}
\caption{
RHEED patterns of a) the Al$_2$O$_3(1\bar{1}$02) substrate,  the Nb buffer layer grown at b) 800$^\circ$C and c) 30$^\circ$C. Growth stages of sample S6 of d) the 6th Fe layer and e) the 6th Nb layer and f)  the protecting Pt cap layer.
}
\label{Fig2}
\end{figure}

The  RHEED pattern of the Al$_2$O$_3(1\bar{1}02)$ substrate (Fig~\ref{Fig2}a) reveals a crystalline structure with Laue rings and  Kikuchi lines  indicating a smooth and ordered surface.
Nb deposition at 800$^\circ$C results in a streaky pattern  and a Laue ring (Fig~\ref{Fig2}b) revealing  epitaxial growth in agreement with previous results \cite{Gute97,Gute97b,wildes2001growth,Nunz96}. In particular, the epitaxial Nb growth of (100) orientation on Al$_2$O$_3(1\bar{1}02)$ substrates was reported in \cite{Nunz96}. The peculiarity of this growth, also seen in our samples, is an $\sim$ 3$^\circ$ angle between the above mentioned planes of Nb and Al$_2$O$_3(1\bar{1}02)$.  At $T_{Nb}=$30$^\circ$C  a transmission pattern  (i.e. a regular arrangement of spots) and rings are visible in the RHEED pattern of the Nb layer which indicate island growth and polycrystallinity, see Fig~\ref{Fig2}c.

Subsequently, the Fe/Nb multilayers were grown on the 800$^\circ$C Nb buffer. The corresponding RHEED patterns exhibit amorphous growth, i.e. blurred screens (not shown). Increasing the Fe film thickness from 2 nm to 4 nm improves the film quality. The Fe layer becomes polycrystalline while the Nb layer remains amorphous. In contrast, for sample S6 which was grown on the 30$^\circ$C Nb buffer, both layers reveal polycrystallinity  with a certain texture, see Figs.~\ref{Fig2}d) and e).
Finally the Pt cap is always  polycrystalline, see Fig.~\ref{Fig2}f).

 \subsubsection{X-ray diffraction}
\begin{figure}[htb]
\centering
\includegraphics[width=1\columnwidth]{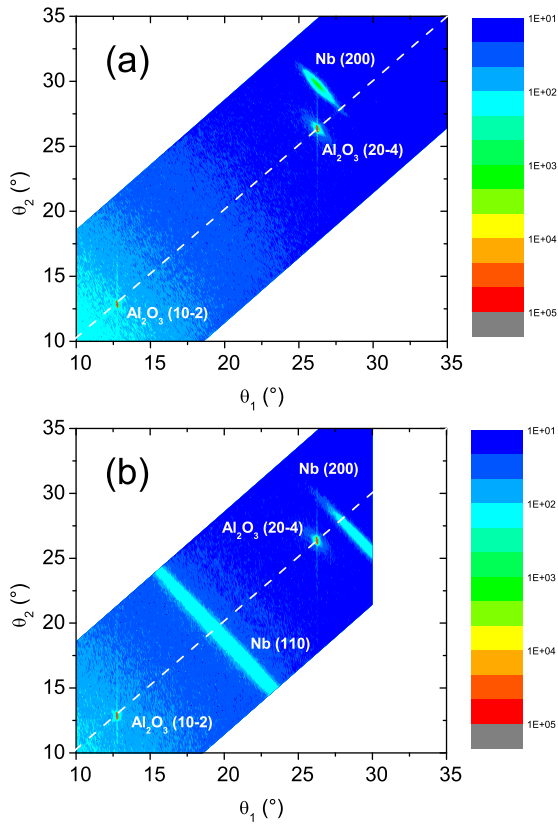}
\caption{
X-ray diffraction patterns for sample s3(a) and s6(b). Dashed tilted lines show the direction of specular reflection. The right bars show the logarithmic intensity scale.}
\label{Fig3}
\end{figure}

 Fig. \ref{Fig3}a shows the diffraction pattern measured on sample s3. Together with two reflections from the substrate we observed a Nb(200) peak at 2$\theta=55^\circ$ with mosaicity of the same order as the substrate peak. In agreement with the observation by RHEED (Fig. \ref{Fig4}b) we observed that the Nb(200) peak is tilted off-specular a couple of degrees which is a well-known feature of Nb growth on Al$_2$O$_3$(1$\bar{1}$02) substrates \cite{Nunz96,wildes2001growth}. Similar patterns were measured for all samples, except for sample s6 which was grown at room temperature. For this sample we measured a typical polycrystalline pattern with coexisting Nb(100) and Nb(110) phases (Fig. \ref{Fig3}b).

 \subsection{Magnetic properties}
 \subsubsection{SQUID measurements}
\begin{figure*}[htb]
\centering
\includegraphics[width=2\columnwidth]{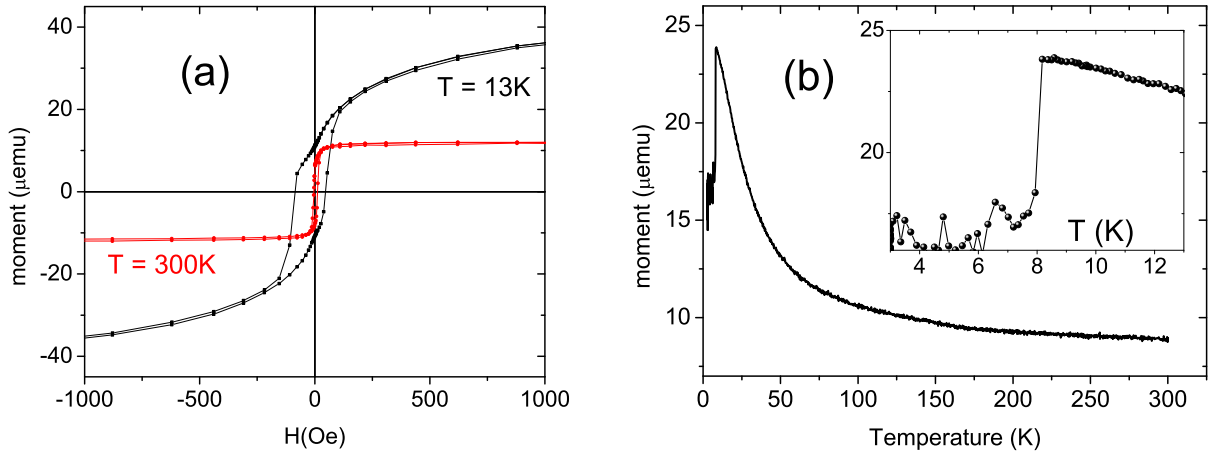}
\caption{
SQUID data of the s3 sample. (a) Hysteresis loop measured at T=300K (red) and T=13K (black). (b) Temperature dependence of magnetic moment measured at H = 250 Oe. The inset shows the data in the vicinity of the superconducting transition.
}
\label{Fig4}
\end{figure*}

 Fig. \ref{Fig4}a shows hysteresis loops measured on sample s3 at $T$ = 300 K and $T$ = 13 K.  At room temperature the sample saturates to a magnetic moment $m_{sat}$ = 12 $\mu$emu above a saturation field of only $H_{sat}$= 50 Oe.  At 13K the saturation moment increases to $m_{sat}$= 40 $\mu$emu and a field above $H_{sat}$ $\approx$ 2 kOe is needed to saturate the magnetic moment of the sample. The temperature dependence of the magnetic moment at $H$ = 250 Oe (Fig. \ref{Fig4}b) shows that the moment is constant down to $T \sim$ 100K, and grows upon further cooling to $T$ = 8.2K. Below this temperature a  downturn of the magnetic moment due to the Meissner effect is observed.

 \subsubsection{Polarized Neutron Reflectometry}
\begin{figure*}[htb]
\centering
\includegraphics[width=2\columnwidth]{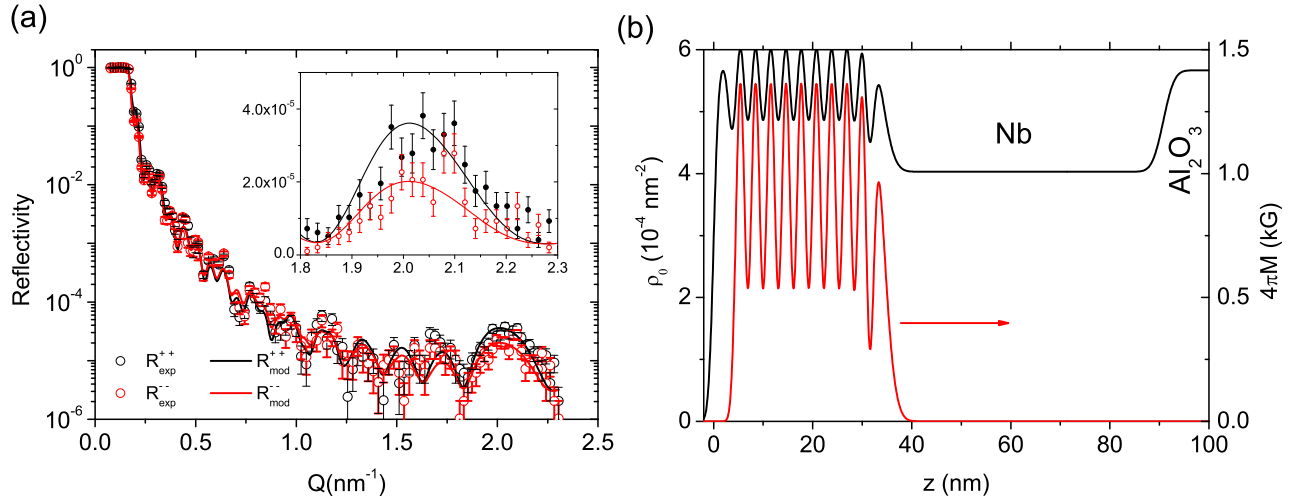}
\caption{
(a) Experimental (dots) and model (lines) reflectivity curves measured on the sample s3 at $T$=13K and $H$=4.5kOe. (b) The depth profiles of SLD and magnetization for the same sample.
}
\label{Fig5}
\end{figure*}

 Fig. \ref{Fig5}a shows reflectivity curves measured on sample s3 at a temperature $T$ = 13K in magnetic field $H$~=~4.5kOe. The curves are characterized by the total external reflection plateau, interference oscillations and the first Bragg peak at $Q_1 \approx $ 2.1 nm$^{-1}$.  The intensity of the Bragg peak $R(Q_1) \equiv R_1  \approx$ 4 $\times$ 10$^{-5}$ is an order of magnitude lower than calculated for the nominal SLDs, indicating high interdiffusion of Fe and Nb. Despite this high interdiffusion we observed a statistically significant difference of Bragg intensities for spin-up and spin-down neutrons (see inset in Fig.\ref{Fig5}a) which suggests the presence of magnetism in the periodic structure. A similar picture was also observed for the samples s1 and s2, which shows that the interdiffusion does not depend strongly on the deposition temperature $T_{PS}$. We fitted experimental curves to models with varying SLDs, thickness, rms roughness of all layers and magnetization of the Fe layer. The resulting depth profiles $\rho_0(z)$ and $M(z)$ are shown in Fig. \ref{Fig5}b. According to our model the SLDs in the center of the Fe and Nb layers is  $\rho_{Fe} = 6.0(2) \times 10^{-4}$ nm$^{-2}$ and $\rho_{Nb} = 5.0(2) \times 10^{-4}$ nm$^{-2}$. Using equation \eqref{Concent} for $N_{av} = (N_{Fe}+N_{Nb})/2$ we can estimate the concentration of iron atoms in the nominal Fe and Nb layers as $c$ = 90\% and $c$=60\%, respectively. In this estimation we used the bulk densities  $N_{Fe}$ = 0.085 \AA$^{-3}$, $N_{Nb}$ = 0.06 \AA$^{-3}$ and scattering lengths $b_{Fe}$ = 9.45 fm and $b_{Nb}$ = 7.05 fm.

 Samples s4 to s6 were measured at room temperature in $H$ = 4.5kOe and analyzed in the same way. The resulting SLDs and magnetization are shown in table \ref{Tab1}. All samples except s4 show strong intermixing of Fe and Nb atoms which resulted in the suppressed magnetization of order of 10\% of the bulk value. For sample s4 with Fe(4nm) the layers become more separated which leads to an increased magnetization of 50\% of the bulk value. Thus neutron reflectometry shows that our periodic structures with $x \le$  2.5 nm form an FeNb alloy with depth-modulated concentration and suppressed magnetization.

  \subsection{Transport measurements}
\begin{figure*}[htb]
\centering
\includegraphics[width=2\columnwidth]{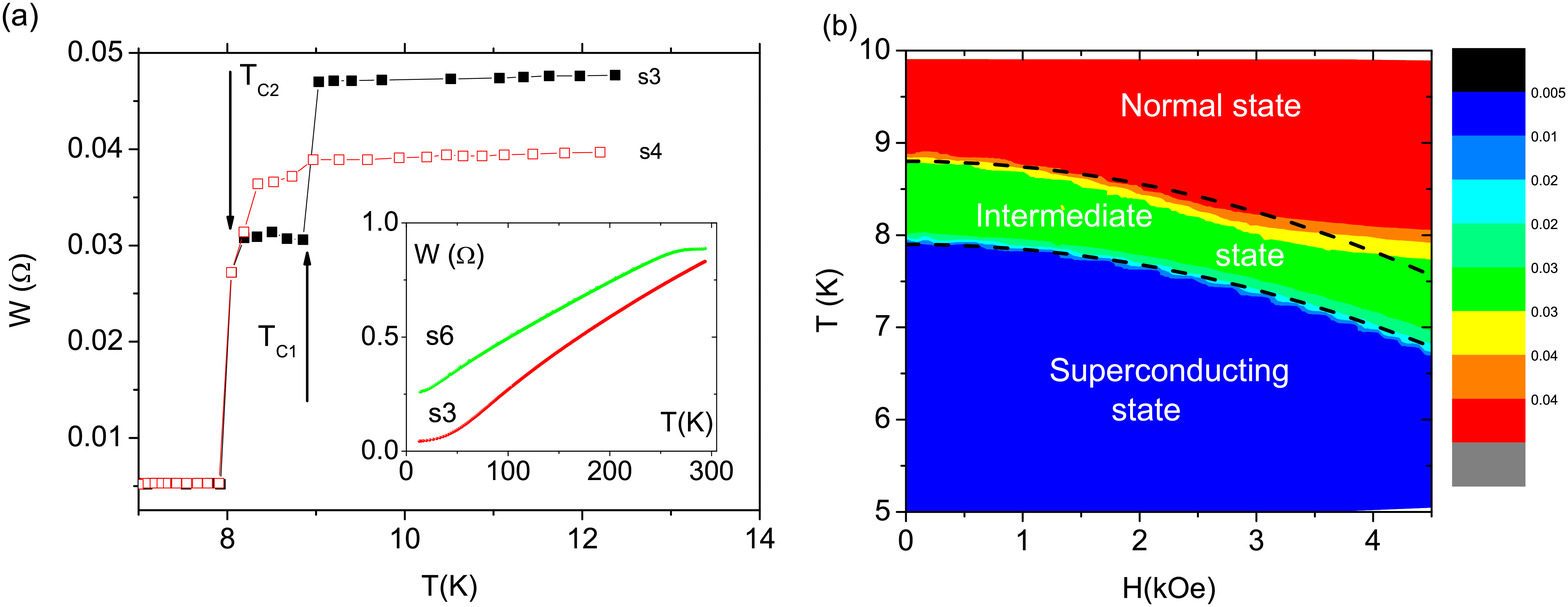}
\caption{
 (a) The $W$($T$) dependence for the samples s4 (black) and s3 (red) in the vicinity of the superconducting transition measured in zero magnetic field. The inset shows the resistance of samples s3 and s6 between room temperature and 10K in $H$ = 4.5kOe. (b) The H-T dependence of resistance of the sample s3. The bottom dashed line show the dependence $T_{c2}(H) = T_{c2}(0) [1-(H/H_{c2}(0))^2] $ with $H_{c2}(0)$ = 12 kOe. The upper dashed line is shifted 0.9K up from the bottom one to show borders of the intermediate state.
}
\label{Fig6}
\end{figure*}

 The inset of Fig. \ref{Fig6}a shows the resistance $W$($T$) of samples s3 and s6 measured during cooling from room temperature to 10K in magnetic field $H$ = 4.5 kOe. For s3 we measured $RRR$ = 18.6, a value which is typical for MBE prepared S/F structures in the epitaxial regime of growth \cite{NowakPRB08,NowakSUST12}. Similar values $RRR$ from 16 to 20 were obtained for all samples except $RRR$ = 3.4 for s6 which was deposited at room temperature and has polycrystalline quality (table \ref{Tab1}).

 Fig. \ref{Fig6}a shows the $W$($T$) curves for the s3 and s4 measured in zero magnetic field in the vicinity of the superconducting transition. For sample s3 we observed a 60\% drop of resistance below $T_{c1}$ = 8.9K. A similar drop was observed for all samples, except for s4, for which the initial drop was only 3\%. Finally, below $T_{c2} \approx$ 8K the resistance falls to zero for all samples, evidencing the superconducting transition. We mention that the second transition coincides with the transition seen by SQUID (inset in Fig. \ref{Fig4}b).

 Fig. \ref{Fig6}b shows the $W$($H$,$T$) phase diagram for sample s3. The superconducting transition can be well described by the expression $T_{c2}(H) = T_{c2}(0) [1-(H/H_{c2}(0))^2] $ with $H_{c2}(0)$ = 12 kOe. This expression can be re-written in the well known form for 2D superconductors: $H_{c2}(T) = H_{c2}(0) \sqrt{1-T/T_{c2}(0)}$. From $H_{c2}(0)$ we can estimate the superconducting correlation length $\xi_S$ = 12 nm, in agreement with the values found in other Nb-based structures \cite{KhaydukovPRB18,KhaydukovPRB19}.

 \section{Results and Discussion}
In this work we studied the structural, magnetic and superconducting properties of [Fe(x)/Nb(1.5)]$_{10}$ superlattices on top of a thick Nb(50) layer.  The main characteristics are collected in table \ref{Tab1}. Our investigation has shown that Nb layer grows epitaxially on the Al$_2$O$_3$(1$\bar{1}$02) substrate in the direction (100) at a temperature of substrate during deposition $T_{Nb}$ = 800$^\circ$C. This result agrees with Ref. \cite{wildes2001growth}. Furthermore the samples grown at this temperature show high residual resistivity ratios of order of 15-20. The sample deposited at room temperature, in contrast, possesses a polycrystalline structure of the Nb(50) layer with a mixture of (100) and (110) phases and  a rather low $RRR$ = 3.4 which is attributed to enhanced scattering of conduction electrons at the grain boundaries.

\begin{table*}[hb]
\caption{\label{tab:Tab1}
Main characteristics of the prepared samples. Here $x$ - thickness of Fe layers, XRD - diffraction peaks observed in XRD experiment, $T_{Nb}$ and $T_{PS}$ are the temperatures of deposition of thick Nb layer and periodic structure,  $\rho_{Fe}$ and $\rho_{Nb}$ are nuclear SLD of in center of nominal Fe and Nb layers, $T_{c1}$ and $T_{c2}$ are the upper and bottom transition temperatures.
}
\begin{ruledtabular}
\begin{tabular}{{cccccccccc}}

ID& x (\AA) & XRD & $T_{Nb}$ ($^\circ$C) & $T_{PS}$ ($^\circ$C) & $\rho_{Fe}/\rho_{Nb}$  & $M_{Fe}$ (kG) & RRR & $T_{c1}$ (K) & $T_{c2}$ (K) \\
\colrule
s1 & 15.4(1) &Nb(200)  & 800  & 100  & 6.1/5.1 & 1(1) & 18.1 & 8.9(3) & 8.1(3) \\
s2 & 15.6(4) &n/m  & 800  & 66  & 5.7/4.7 & 1.8(8) & 16.9 & 8.9(1) & 8.0(2) \\
s3 & 15.1(1) &Nb(200)  & 800  & 30  & 6.0/4.9 & 2(1) & 18.6 & 8.8(2) & 8.1(2) \\
s4 & 36.7(3) &Nb(200)  & 800  & 30  & 6.6/4.0 & 11.2(4) & 19.6 & 8.2(2) & 8.0(1) \\
s5 & 26.7(4) &Nb(200)  & 800  & 30  & 6.1/3.9 & 2(1) & 16.3 & 8.9(1) & 7.9(2) \\
s6 & 20.1(5) &Nb(110),Nb(200)  & 30  & 30  & 5.0/4.6 & 2.3(5) & 3.4 & 9.1(1) & 8.3(1) \\

\end{tabular}
\end{ruledtabular}
\label{Tab1}
\end{table*}

Neutron reflectometry has shown that Fe/Nb superlattices with $x \le$ 2.5 nm form a depth-modulated FeNb alloy with concentration of iron varying within the superlattice unit cell between 90\% and 60\%. Based on the SQUID data (Fig. \ref{Fig4}) we can attribute the magnetic signal at room temperature to the iron-rich Fe$_{0.9}$Nb$_{0.1}$ alloy, while the signal below $T_m \sim$ 100 K originates from Fe$_{0.6}$Nb$_{0.4}$. Though the thickness of our Nb spacer, 1.3nm, is close to the values in Refs. \cite{RehmPhysB96,RehmEPL97,KlosePRL97} we did not observe any antiferromagnetic coupling, neither at room temperature nor in low-temperature measurements. The reason of this disagreement may originate from the amorphous Nb spacers.
The proximity of this depth modulated and weakly magnetic layer to a thick superconductor causes the appearance of an intermediate phase between the normal state ($T > T_{c1}$) with nonzero resistance and the superconducting state ($T < T_{c2}$) with zero resistance. This state is characterized by a $\sim$50\% suppressed resistance and absence of the Meissner effect. We note that a stand-alone Fe/Nb periodic structure itself can not be a superconductor due to the absence of a clean and oriented Nb phase. In this regard we may conclude that the intermediate state is a result of the proximity effect to the thick Nb layer. This proximity leads to the appearance of superconducting correlations in the periodic structure. However, close to $T_C$, the density of superconducting correlations is not enough to form a homogeneous superconducting phase in the whole structure. Thus an inhomogeneous state with mixture of superconducting and normal-state domains both in the Nb(50nm) buffer layer and Fe/Nb superlattice is formed leading to suppressed (but nonvanishing) resistance and absence of the Meissner effect.

In conclusion, we studied the structural, magnetic and superconducting properties of [Nb(1.5)/Fe(x)]$_{10}$ superlattices deposited on a thick Nb(50) layer.  Our investigation showed that the high deposition temperature $T_{Nb}$ = 800$^\circ$C results in high structural quality systems with epitaxial Nb(50nm) layer and high residual resistivity ratio of order of 15-20. By using neutron reflectometry we have shown that Fe/Nb superlattices with $x<$ 4 nm form a depth-modulated FeNb alloy with concentration of iron varying between 90\% and 60\%. This alloy has properties of a weak ferromagnet with Curie temperature of order of $T_m \sim$ 100K. Proximity of this weak F layer to a thick superconductor leads to the presence of an intermediate phase between normal and superconducting state. This phase is characterized by co-existed domains of superconducting and normal-state phases. By increasing thickness of Fe layer to $x$ = 4 nm this phase was destroyed.

\begin{acknowledgments}
We would like to thank G. Logvenov, F. Klose, and Ch. Rehm  for the fruitful discussions.  YK, TK and BK would like to acknowledge financial support of German Research Foundation (Deutsche Forschungsgemeinschaft, DFG, Project No. 107745057 - TRR80). This work is partially based on experiments performed at the NREX instrument operated by Max-Planck Society at the Heinz Maier-Leibnitz Zentrum (MLZ), Garching, Germany. Sample preparation was performed in the thin film laboratory of the J\"ulich Centre for Neutron Science (JCNS) at Heinz Maier-Leibnitz Zentrum (MLZ), Garching, Germany. AS and RM would like to thank the support of the "SPINTECH" project of the HORIZON-2020 TWINNING program (2018-2020).
\end{acknowledgments}

\bibliography{FeNb_Refs}

\end{document}